\documentclass[runningheads]{llncs}

\usepackage[T1]{fontenc}
\usepackage{xcolor}
\usepackage{graphicx}
\usepackage{amsmath}
\usepackage{graphicx} 
\usepackage{float}
\usepackage[utf8]{inputenc}
\usepackage{url}        
\usepackage{hyperref}   
\usepackage{cite}

\newif\ifpreprint
\preprinttrue            

\usepackage[T1]{fontenc}
\usepackage[utf8]{inputenc}
\usepackage{xcolor}
\usepackage{graphicx}    
\usepackage{amsmath}
\usepackage{float}
\usepackage{url}
\usepackage{cite}
\usepackage{hyperref}    

\ifpreprint
  \usepackage{draftwatermark}
  \SetWatermarkText{Preprint}
  \SetWatermarkScale{0.17}
  \SetWatermarkColor[gray]{0.90}
\fi

\ifpreprint
  \hypersetup{
    pdfauthor={Jingwen Qin, Semen Checherin, Yue Li, Berend-Jan van der Zwaag, Özlem Durmaz-Incel},
    pdftitle={Hue4U: Real-Time Personalized Color Correction in Augmented Reality (Preprint)},
    pdfsubject={Preprint on arXiv},
    pdfkeywords={Color Vision Deficiency, Augmented Reality, Personalized Color Correction, Preprint}
  }
\fi

\begin{document}

\title{Hue4U: Real-Time Personalized Color Correction in Augmented Reality}
%
%
\author{Jingwen Qin\inst{1}\orcidID{0009-0003-2421-170X} \and
Semen Checherin\inst{1}\orcidID{0009-0003-8795-9563
 } \and
Yue Li\inst{2}\orcidID{0009-0001-2375-5997} \and
Berend-Jan van der Zwaag\inst{1}\orcidID{0000-0003-0884-5334
} \and
Özlem Durmaz - Incel\inst{1}\orcidID{0000-0002-6229-7343}
}

\authorrunning{J. Qin et al.}

\institute{
Faculty of EEMCS, University of Twente, Enschede, The Netherlands \and
Faculty of BMS, University of Twente, Enschede, The Netherlands \\
\email{gwen.qin@utwente.nl}
}

\maketitle              

\ifpreprint
\begin{center}
  \vspace{-0.5em}
  {\small
  \textbf{Preprint (arXiv), version \today.}\\
  Accepted at \emph{IWOAR 2025}.\\
  This is the authors' preprint; the final version will appear in the Springer proceedings.\\
  If citing, please use the arXiv identifier: \href{https://arxiv.org/abs/2509.06776}{arXiv:2509.06776}.
  }
\end{center}
\vspace{0.5em}
\fi

\begin{abstract}
Color Vision Deficiency (CVD) affects nearly 8\% of men and 0.5\% of women worldwide. Existing color correction methods often rely on prior clinical diagnosis and static filtering, making them less effective for users with mild or moderate CVD. In this paper, we introduce Hue4U, a personalized, real-time color correction system in Augmented Reality using consumer-grade Meta Quest headsets. Unlike previous methods, our Hue4U requires no prior medical diagnosis and adapts to the user in real time. A user study with 10 participants showed notable improvements in their ability to distinguish colours. The results demonstrated large effect sizes (Cohen’s $d > 1.4$), suggesting clinically meaningful gains for individuals with CVD. These findings highlight the potential of personalized AR interventions to improve visual accessibility and quality of life for people affected by CVD.

\end{abstract}
\section{Introduction}
Color Vision Deficiency (CVD) is a common, yet often overlooked, sensory condition that affects approximately 8\% of men and 0.5\% of women globally~\cite{Simunovic2009}. It is caused by anomalies in one or more types of cone photo-receptors (L-cones, M-cones, and S-cones) in the retina, leading to varying degrees of difficulty in distinguishing specific hues. The severity of symptoms can vary widely between individuals. There are two major types of color vision deficiency (CVD): dichromacy, in which one of the three cone types is completely non-functional, making certain colors impossible to perceive; and anomalous trichromacy, where all three cone types are present but one functions abnormally. These deficiencies are further categorized as protan, deutan, and tritan. Although CVD is not life-threatening, it can significantly impact daily life, especially in tasks where color discrimination is essential, for example, selecting food products, cooking, playing sports, or choosing clothing. It can also limit professional opportunities in fields such as aviation, engineering, graphic design, and certain military roles. Furthermore, difficulties in interpreting color-coded information can lead to frustration and social challenges, ultimately reducing the overall quality of life \cite{Melillo2017}. Currently, there is no medical cure for CVD. 

Several digital and optical tools are used to address these challenges, including Daltonization filters\cite{Vienot1999}, contrast-enhancing overlays, and wearable color correction lenses. For example, Tang \cite{Tang2020} developed AR-based contrast-enhancing overlays using Daltonization to assist red–green CVD, while Tian \cite{Tian2022} introduced inverse-designed contact lenses that optically shift colors to match individual photoreceptor sensitivities. However, most of these methods require users to first obtain a clinical diagnosis of their CVD type, even though many individuals are undiagnosed and unaware of their condition. In addition, these solutions typically apply static transformations optimized for severe dichromacy, neglecting the more common cases of mild or moderate anomalous trichromacy. As a result, such "one-size-fits-all" approaches fail to address the broad spectrum of CVD severity and provide only limited solutions to most users.

Augmented Reality (AR) technologies offer a promising way to enhance visual accessibility with more interactive methods. AR, typically delivered through Head-Mounted Displays (HMDs), enables users to remain fully aware of their physical surroundings while overlaying digital content onto the real world. This makes AR particularly well-suited for color correction tasks, such as identifying food, clothing, or traffic signals in daily life. In addition, AR platforms can support personalized color filtering without interrupting natural interactions. 

This work addresses the research question of how to design a real-time, personalized color correction system in AR that improves color discrimination for individuals with CVD without prior clinical diagnosis. To solve this issue, we present an AR-based personalized color correction, named Hue4U, to develop a personalized color correction system for CVD users, including on-device CVD estimation and real-time color transformation. Our Hue4U system consists of two key components: (1) an interactive color vision assessment module integrated into an AR environment, and (2) a real-time color correction module that applies user-specific transformations based on the estimated CVD type and severity. The novelty of this work lies in the fact that the entire pipeline operates within AR, requires no prior clinical diagnosis, and adapts to the user’s individual color perception profile. Another important aspect is the implementation and evaluation of the Hue4U system on modern, consumer-grade, and widely accessible Meta Quest. We recruited 10 participants with varying degrees and types of CVD. Each participant first completed a standardized assessment to determine their specific CVD profile. Based on these results, a personalized color correction filter was applied for real-time rendering in the AR environment.  We evaluate our Hue4U system using quantitative metrics, including pre- and post-correction performance on Ishihara plate tests \cite{Ishihara1917}. Our results show consistent improvements in color discrimination, with most participants able to identify significantly more plates after correction. We make the following key contributions:

\begin{itemize}
    \item We propose Hue4U, a real-time, personalized color correction system in AR that improves color discrimination for individuals with CVD without requiring prior clinical diagnosis.

    \item We design an adapted FM 100 test on HMD with hand controllers in mixed lighting environments, considering user fatigue and display limitations.

   \item We apply an existing Daltonization method in a novel way by integrating severity-adaptive, per-pixel color correction into Unity’s shaders, enabling personalized real-time recoloring for AR content based on each user’s CVD profile.

    \item We conduct a user study with 10 participants, showing statistically significant improvements in Ishihara test performance after color correction, with large effect sizes (Cohen’s $d > 1.4$).

\end{itemize}

The remainder of this paper is organized as follows: Section \ref{sec:relatedwork} reviews related work on CVD assessment, color simulation, and recoloring methods. Section \ref{sec:methodology} details our methodology, including the Hue4U system architecture, CVD assessment, recoloring, and rendering techniques. Section \ref{sec:experimental setup} describes the experimental setup, including hardware and development environment. Section \ref{sec:quantitative assessment of filter efficacy} presents the quantitative assessment procedure and evaluation of filter efficacy. Finally, Section \ref{sec:conclusion} concludes the paper and outlines directions for future work.

\section{Related Work}
\label{sec:relatedwork}
\subsection{CVD Assessment}
A CVD test is a diagnostic tool used to identify the type and severity of color vision deficiency. In our Hue4U system, we use CVD tests to establish the ground truth—determining whether individuals are color blind and assessing the severity of their symptoms. It is essential since most people have never been formally diagnosed and therefore do not know if they have CVD or how severe their condition is. Additionally, after applying our real-time color correction module, we use CVD tests again to evaluate how effectively users can perceive colors.

\begin{figure}[t!]
    \centering
    \includegraphics[width=0.4\linewidth]{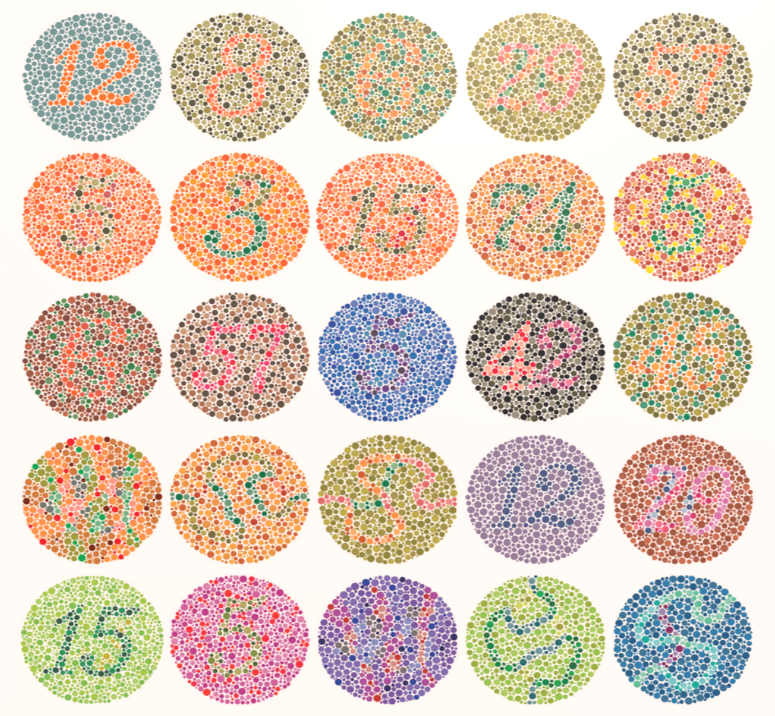}  
    \caption{Sample Ishihara Plates Used in the Study: Examples of the Ishihara plates presented to participants during the color discrimination tests. A total of 25 plates were used to evaluate the participants’ CVD both before and after applying personalized color correction in the AR environment.}
    \label{fig:Ishihara-Plate}
\end{figure}

The most popular test is the Ishihara plate test \cite{Ishihara1917}, first developed in 1917 to detect deficiencies in the medium- and long-wavelength (M-L) cones. The test consists of several plates, each displaying a circle made up of dots in varying colors, brightnesses, and sizes, shown in Figure~\ref{fig:Ishihara-Plate}. Embedded within these dots is a number or path formed by slightly different colors. Individuals with CVD may either fail to see the number or path or may perceive a different figure altogether. The Ishihara test is simple, quick, non-invasive, and easy to administer. However, it does not effectively quantify the severity of color vision deficiency. Moreover, the Ishihara test primarily targets red-green deficiencies, it was not made to estimate tritanopia (a complete absence of S-cones in the eye). Additionally, individuals with acquired CVD may have unusual or distorted cone sensitivities, causing them to perceive colors differently than expected\cite{Simunovic2016}. Such individuals might still pass the Ishihara test despite having a color vision problem. 
Another widely used test is the Farnsworth-Munsell 100-Hue test (FM 100 test), designed by Farnsworth\cite{verriest_new_1982}. This test includes 85 colored caps distributed across four trays, with hues gradually changing within each tray. Participants are asked to arrange the caps in order of hue progression to create a smooth color transition. The FM 100 test is highly recommended for detecting and diagnosing acquired color vision deficiencies. Unlike the Ishihara test, which primarily screens for red-green CVD, the FM 100 test can detect all types of color vision deficiencies and assess how severe their deficiencies are. There are also simplified versions of the FM 100 test, such as the D15 test, which uses only 15 colored caps\cite{Arpadffy-Lovas2024}. However, these shorter tests are less sensitive to mild color vision deficiencies and do not provide detailed information on the severity of the condition.

\subsection{CVD Color Simulation}

Individuals with CVD perceive colors differently from those with normal vision. Researchers have developed image recoloring techniques that adjust the colors in images to enhance color distinguishability. In our Hue4U system, we employ the color simulation method to simulate how an image would appear to a person with CVD. We then use this simulation to compute the color discrepancy between the original image and its CVD-simulated counterpart. This allows us to quantify potential perceptual errors and design targeted color correction strategies.

Brettel \cite{brettel_computerized_1997} developed one of the earliest and most influential computerized simulations of CVD. Their method converts images from the RGB color space to the LMS color space, which models the response of the human eye’s three types of cones (long-, medium-, and short-wavelength). It then simulates the absence of one cone type by projecting the color information onto a confusion line based on the remaining cones\cite{brettel_computerized_1997}. In this way, it simulates how images appear to individuals with dichromacy. However, it does not simulate anomalous trichromacy, where all three cones are present but function abnormally.

Machado \cite{machado_physiologically-based_2009} proposed a more advanced and physiologically accurate simulation. Their model first converts image colors into a physiologically relevant color space, then applies severity parameters to adjust the colors, and finally transforms the result back into the RGB color space\cite{machado_physiologically-based_2009}. This method supports both dichromacy and anomalous trichromacy and allows for the simulation of varying degrees of color vision deficiency through the use of severity parameters. Additionally, since Machado’s model has been widely implemented in existing software frameworks, it is practical, accessible, and easy to integrate into applications.

\subsection{CVD Recoloring}
Individuals with CVD have difficulty distinguishing certain colors. Color correction techniques aim to enhance the contrast and visibility of these indistinguishable colors without making the overall image appear unnatural. Among the most widely used correction methods are Daltonization \cite{Dalton1794} and Natural-Preserving Recoloring \cite{zhou_fast_2024}. 

Daltonization first simulates how an image appears to a CVD person using color vision simulation techniques. The system compares the original and simulated images to detect and restore lost color contrasts\cite{Dalton1794}. Due to its computational efficiency and speed, Daltonization is well-suited for real-time applications. However, most Daltonization algorithms are based on dichromacy and rely on fixed color mappings. As a result, they do not account for the varying severity of CVD and may produce visually unnatural results for users with anomalous trichromacy.

To address this limitation, Zhu \cite{Zhu2022} proposed a Natural-Preserving Recoloring algorithm, which improves color distinguishability while maintaining the overall natural appearance of the image. This method first simulates CVD perception and then optimizes color adjustments to balance enhanced contrast with visual fidelity\cite{Zhu2022}. Although this approach is more visually subtle and user-friendly—particularly for individuals with anomalous trichromacy—it is computationally expensive and unsuitable for real-time systems, requiring approximately thirty seconds to process a single image. Given the real-time requirements of AR systems, we consider Daltonization to be the more feasible approach for practical deployment.

\section{Methodology}
\label{sec:methodology}

\subsection{System Architecture}

\begin{figure}[H]
    \centering
    \includegraphics[width=1\linewidth]{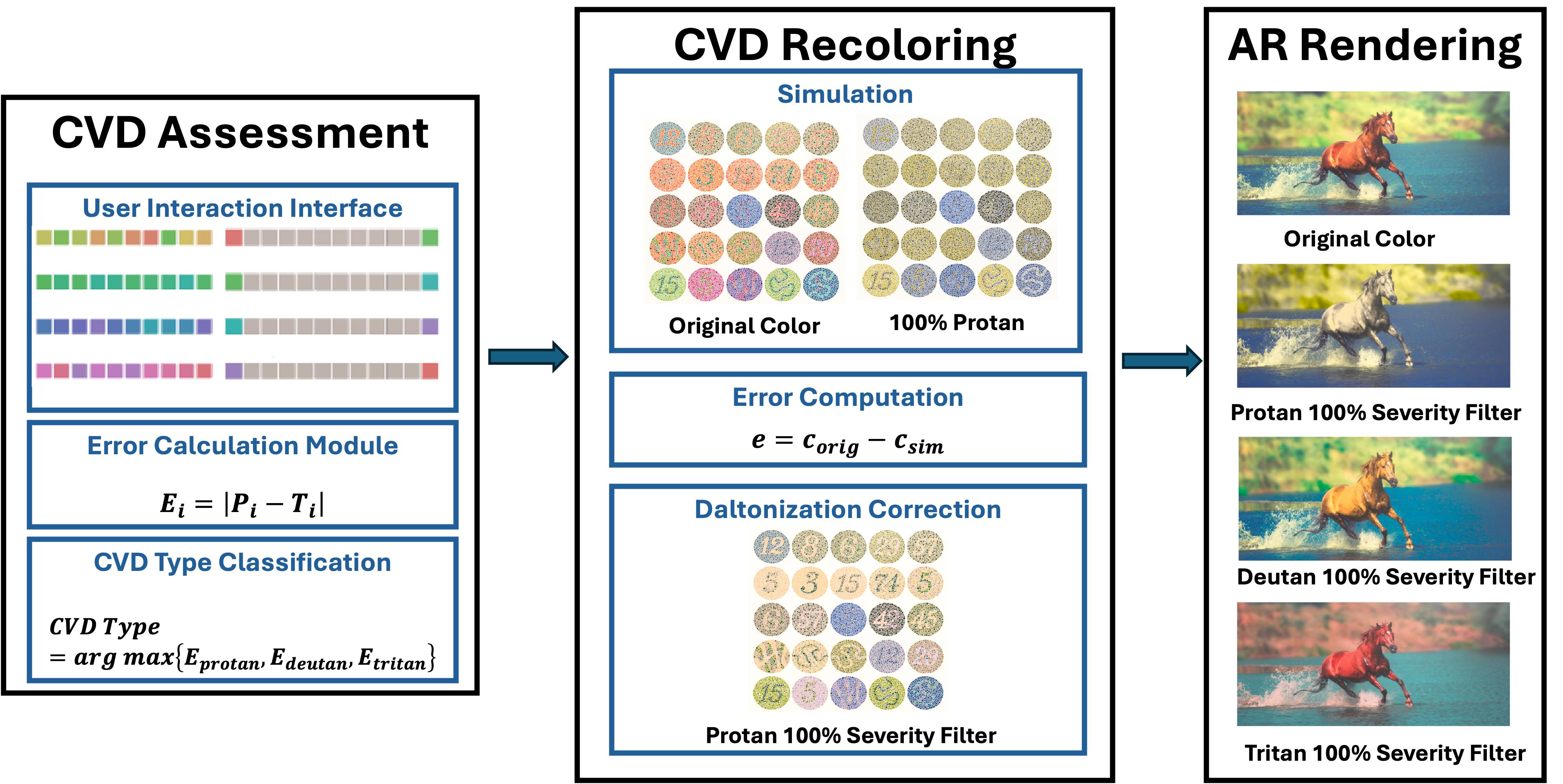}  
    \caption{Hue4U: End-to-End Pipeline for CVD Assessment and Real-Time Color Correction in AR: The workflow illustrates how to use our adapted FM 100 test results for personalized color correction for AR elements. Based on the user's rearrangement of 48 hue caps, the Hue4U system computes positional errors to classify the CVD type and estimate severity (see Section 3.2). This personalized CVD profile is passed to a correction controller, which selects the appropriate Daltonization parameters (Machado matrix) and forwards them to a custom Unity shader(see Section 3.3). The shader then performs real-time per-pixel color correction on all AR content, ensuring improved color discriminability for users with CVD (see Section 3.4).}
    \label{fig:system-pipeline}
\end{figure}

Hue4U is an end-to-end system for CVD assessment and personalized visual correction in a Unity-based AR environment. It translates user input from an integrated color vision test into personalized color correction for virtual AR content. It consists of three primary components shown in Figure~\ref{fig:system-pipeline}. The CVD Assessment module implements a modified version of the FM 100 Test to evaluate both the type and severity of CVD. Users interact with colored tiles that must be reordered to form perceptual hue gradients. Based on the arrangement errors, the Hue4U system computes a cumulative error score and classifies the CVD type and severity. The CVD Recoloring module applies personalized color correction based on the results of the CVD assessment. Utilizing custom real-time shaders, it performs personalized Daltonization to enhance color discriminability according to the user's specific CVD profile. AR Rendering integrates color correction into Unity’s Universal Render Pipeline (URP) via custom pixel shaders. These shaders apply per-pixel Daltonization in real time, adjusting the color output of virtual objects based on the user’s CVD profile.

\subsection{CVD Assessment}
Since the FM 100 test can detect all types of CVD and assess their severity, we selected this method for CVD assessment in our Hue4U system.

\subsubsection*{Adapted FM 100 Test.}
The original FM 100 test requires the arrangement of 85 color caps distributed across four trays. However, directly implementing this test in an AR HMD setup presents several challenges. Although the Meta Quest 3 HMD offers high resolution and improved color contrast and brightness compared to earlier versions, it uses LCD panels with a limited color gamut. As a result, some subtle hues may not be rendered accurately. Displaying all 85 hues can lead to indistinguishable colors, particularly those near the limits of the device's display capabilities. Moreover, completing the full 85-cap test in AR is both cognitively and physically demanding for users when using handheld controllers. To reduce user fatigue while maintaining diagnostic effectiveness, we opted for a shorter version of the test. However, reducing the number of caps to 48 does not compromise our test’s accuracy, as the diagnostic outcome is not based on the number of caps. Therefore, our adapted version still provides reliable identification of CVD type and severity.

Additionally, ambient lighting and background colors in the physical environment can influence the perceived colors of AR elements in HMD. To reduce these effects, we introduced a virtual white board behind the test interface. In this way, it minimizes changes in contrast due to lighting, eliminates background color interference, and ensures a more consistent and visually reliable user experience.

We conducted several pilot tests with participants with normal color vision and asked them to report perceived color differences. Based on these results, we selected 48 caps instead of the full 85. Increasing the number of caps beyond this point introduced difficulty in distinguishing between adjacent hues. 

Our revised FM 100 test is implemented in two formats: one version runs on the HMD for our Hue4U system, and the other on a standard laptop to obtain the ground truth CVD profile. In the HMD version, users interact with the caps using handheld controllers, while in the desktop version, they use a mouse. Same as in the original test, the first and last caps of each row are fixed, and users must rearrange the intermediate caps to form a smooth hue transition from left to right. The interface includes visual and haptic feedback and automatically aligns the selection ray toward the nearest cap. In this way, it enhances usability, intuitiveness, and overall user experience.

\subsubsection*{Design and Implementation.}
\begin{figure}[htbp]
    \centering
    \includegraphics[width=0.5\linewidth]{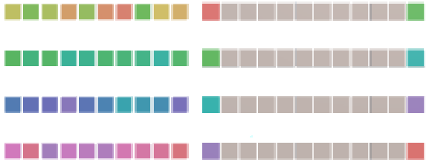}  
    \caption{Adapted FM100 Test in AR Environment: Users drag colorful caps from the randomized collection on the left side of the interface to the corresponding grey slots on the right, arranging them to form a smooth hue gradient. The first and last squares in each row of the right side of the interface are fixed reference colors that define the gradient boundaries. }
    \label{fig:FM_100_Test}
\end{figure}

Our adapted test comprises four rows. And same as the original FM 100 test, the first and last caps of each row are fixed reference points that define the boundaries of the hue gradient. However, instead of 21 movable caps per row, our version employs 11 movable caps per row, resulting in a total of 48 movable caps across all four rows. Each row represents a hue gradient between two predefined anchor colors. Table~\ref{tab1} summarizes the hue transitions and their corresponding diagnostic targets across the four rows of our adapted FM 100 test. Figure~\ref{fig:FM_100_Test} illustrates the implementation of the adapted FM 100 test in an AR environment. Each row represents a controlled gradient between two anchor colors selected to evaluate specific types of CVD. These anchors were carefully selected to span hue transitions most relevant for detecting the three major types of color vision deficiencies. The RGB values of the four anchor colors used in our system are the same as the original FM 100 test:
\begin{itemize}
    \item {Red}: \texttt{(0.75, 0.40, 0.40)}
    \item {Yellow-Green}: \texttt{(0.30, 0.67, 0.33)}
    \item {Cyan-Green}: \texttt{(0.10, 0.65, 0.65)}
    \item {Blue-Purple}: \texttt{(0.52, 0.46, 0.71)}
\end{itemize}

To generate perceptually accurate intermediate hues between these anchors, we interpolate the colors in the Hue-Saturation-Value (HSV) color space rather than in standard RGB. Since RGB interpolation often results in variations in saturation, it can compromise the perceptual consistency required for accurate color vision testing. However, HSV space maintains constant saturation and brightness to ensure only hue varies throughout each gradient. Once the interpolation is complete, the resulting values are converted back to RGB for rendering in AR environment.

\begin{table}
\caption{Hue Ranges and Corresponding CVD Types in the Adapted FM 100 Test.}
\label{tab1}
\begin{tabular}{|l|l|l|l|}
\hline
\textbf{Row} & \textbf{Caps Range} & \textbf{Hue Range} & \textbf{Related CVD Types} \\
\hline
1 & 1--12   & Red $\rightarrow$ Yellow-Green           & Protan, Deutan \\
2 & 13--24  & Yellow-Green $\rightarrow$ Cyan-Green    & Deutan \\
3 & 25--36  & Cyan-Green $\rightarrow$ Blue-Purple     & Tritan \\
4 & 37--48  & Blue-Purple $\rightarrow$ Red & Protan, Tritan \\
\hline
\end{tabular}
\end{table}

\subsubsection*{Cap Error and Total Error Score.}
Each of the four rows in our test contains $n = 11$ movable hue caps. For each cap $i$, let $P_i$ denote the \emph{actual position} of the cap after user reordering, and $T_i$ be the \emph{target position} in the correct hue gradient. 
The \textbf{cap error} is defined as:

\begin{equation}
E_i = |P_i - T_i|
\end{equation}

The total error for one row is the sum of individual cap errors:

\begin{equation}
E_{\text{row}} = \sum_{i=1}^{n} E_i
\end{equation}

The total error across all four rows is calculated as:

\begin{equation}
E_{\text{total}} = \sum_{r=1}^{4} E_{\text{row}_r}
\end{equation}
To make the error score comparable to the original FM 100 test, which uses 85 movable caps, we normalize the error by scaling it:

\begin{equation}
E_{\text{scaled}} = \frac{E_{\text{total}}}{48} \times 85
\end{equation}

This scaled error score serves as an estimate of the severity of the user's color vision deficiency.

\subsubsection*{CVD Type Classification.}
Accordingly, we calculate cumulative error scores associated with each CVD type:

\begin{align}
E_{\text{protan}} &= E_{\text{row}_1} + E_{\text{row}_4} \\
E_{\text{deutan}} &= E_{\text{row}_1} + E_{\text{row}_2} \\
E_{\text{tritan}} &= E_{\text{row}_3} + E_{\text{row}_4}
\end{align}

The CVD type is classified based on the highest cumulative error:

\begin{equation}
\text{CVD Type} = \arg\max \left\{ E_{\text{protan}}, E_{\text{deutan}}, E_{\text{tritan}} \right\}
\end{equation}

This error-based classification allows our shortened FM 100 test to effectively detect and differentiate between protan, deutan, and tritan color vision deficiencies, while also estimating CVD severity in a scaled and interpretable manner.

\subsection{CVD Recoloring}
\begin{figure}[H]
    \centering
    \includegraphics[width=1\linewidth]{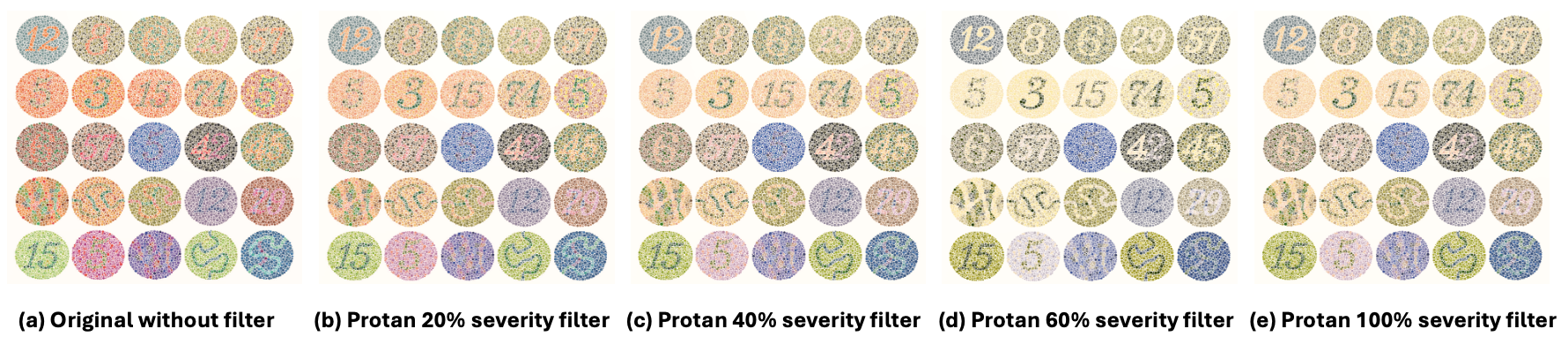}  
    \caption{Visualization of the 25 Ishihara Plate after CVD recoloring: This figure illustrates the visual appearance of the 25 Ishihara plate as perceived by individuals with varying severity of protan CVD. From left to right, the images display the original plate, followed by protan recoloring filters applied at 20\%, 40\%, 60\%, and 100\% severity levels.}
    \label{fig:Method_Results}
\end{figure}

To enhance accessibility for users with CVD, we adopt a Daltonization-based color correction strategy. Our method follows a standard three-stage pipeline: simulation, error computation, and correction. By integrating a parameterized model of color perception loss, our Hue4U system enables real-time, personalized recoloring based on an individual’s CVD type and severity level. Figure~\ref{fig:Method_Results} demonstrates after 3 stages of our CVD recoloring method, how color perception progressively changes with increasing severity of protan deficiency.

\paragraph{Simulation}

In the first stage, we simulate how colors appear to individuals with a specific type and severity of CVD. The method proposed by Machado \cite{machado_physiologically-based_2009} provides a set of transformation matrices that transform between normal and CVD individuals. These matrices are parameterized by a severity value \( s \in [0, 1] \), where \( s = 0 \) represents normal color vision and \( s = 1 \) corresponds to complete color blindness. Given an input color vector
\begin{equation}
\mathbf{c}_{\text{orig}} = 
\begin{bmatrix}
R \\
G \\
B
\end{bmatrix},
\end{equation}
the simulated color \( \mathbf{c}_{\text{sim}} \) for a given CVD type and severity level is calculated as:
\begin{equation}
\mathbf{c}_{\text{sim}} = M_{\text{CVD}}(s) \cdot \mathbf{c}_{\text{orig}},
\end{equation}
where \( M_{\text{CVD}}(s) \) is the transformation matrix corresponding to the specified CVD type and severity.

\paragraph{Error Computation}
The second stage involves calculating the difference between the original color and its corresponding CVD simulation. This difference represents the color information lost due to CVD:

\begin{equation}
\mathbf{e} = \mathbf{c}_{\text{orig}} - \mathbf{c}_{\text{sim}}.
\end{equation}

\paragraph{Daltonization Correction}
In the final stage, the error is redistributed across visible color channels via a predefined correction matrix \( C_{\text{type}} \). The corrected color \( \mathbf{c}_{\text{corr}} \) is calculated as:  
\begin{equation}
\mathbf{c}_{\text{corr}} = \mathbf{c}_{\text{orig}} + C_{\text{type}} \cdot \mathbf{e}
\end{equation}

This process shifts the original color so that it enhances its perceptual distinguishability for the CVD viewer, without introducing significant visual artifacts for individuals with normal color vision.

\subsection{AR Rendering}

\begin{figure}[htbp]
    \centering
    \includegraphics[width=1\linewidth]{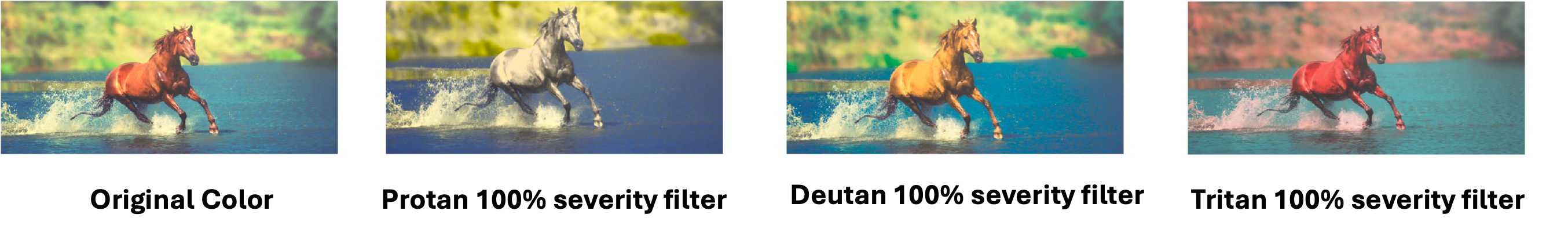}  
    \caption{AR rendering output using our Hue4U system: From left to right: (1) Original color perception in AR; (2) Protan (100\% severity); (3) Deutan (100\% severity); (4) Tritan (100\% severity).  }
    \label{fig:recolor}
\end{figure}

Our Hue4U system renders virtual objects in AR using URP to ensure real-time performance and accuracy. We apply Daltonization on a per-pixel basis in URP. This enables continuous and fine-grained correction adjustments directly within the scene, using high-precision floating-point color computations. This shader independently executes the three-stage process described in Section 3.3. Simulation matrices for each CVD type and severity are stored within a separate class and accessed by a correction controller at runtime. Based on the user's estimated CVD profile, the controller loads the corresponding Machado matrix. The shader then computes and applies the Daltonization correction in real-time, preserving critical color characteristics for CVD users. Due to the simplicity of our AR scene, per-pixel correction remains computationally efficient while maintaining smooth performance on Meta Quest 3. Figure~\ref{fig:recolor} illustrates the AR rendering results, including the original AR view and various filtered renderings.

\section{Experimental Setup}
\label{sec:experimental setup}

\subsection{Implementation and Experimental Setup}
Our Hue4U system is developed using Unity 2022.3 LTS and URP as a standard for rendering in AR applications. To ensure broad compatibility across various HMDs, we integrate OpenXR \cite{OpenXR} for control and input management. Additionally, we utilize MetaSDK 77, which provides streamlined access to pass-through APIs, enabling precise and efficient color correction.

Among the many available HMD options, we selected the Meta Quest 3 \cite{MetaQuest3} due to its optimal balance of accessibility, technical capability, and user experience. The device features high-resolution LCD panels with enhanced color contrast and brightness. It is essential for accurately rendering subtle hue variations necessary for effective CVD assessment and correction. Furthermore, as a consumer-grade and widely accessible device, the Meta Quest 3 maximizes the social impact of our Hue4U system by delivering personalized CVD correction to a broad audience without the need for expensive hardware or prior clinical diagnosis. We log the output of our system during the experiments, and the results show that with the filter, the performance of the system is around 150ms.

Our participant pool consisted of 10 individuals, primarily exhibiting protan and deutan characteristics, which together account for over 95\% of CVD cases. Although most participants were not clinically diagnosed, initial screenings using the FM 100 test and Ishihara plates revealed deviations from normal color vision. The baseline FM 100 assessment further provided quantitative estimates of each participant’s CVD type and severity. All procedures were reviewed and approved by the CIS Ethics Committee at University of Twente, and informed consent was obtained from all participants prior to the study.

\subsection{Experiment Procedure}
\begin{figure}[H]
    \centering
    \includegraphics[width=1\linewidth]{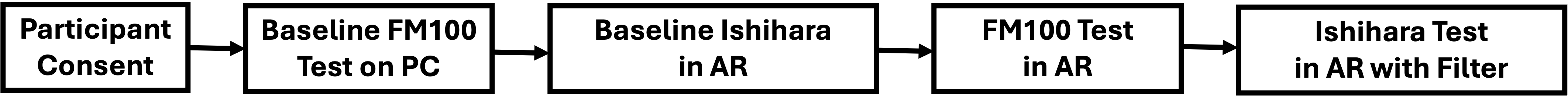}
    \caption{Overview of the Experimental Procedure: Flowchart illustrating the tasks performed by participants during the study. The procedure includes the adopted FM 100 Test on PC for ground truth measurement, baseline Ishihara plate test in AR without color correction, the adapted FM100 test in AR used for personalized recoloring, and the final Ishihara plate test with color correction.}
    \label{fig:Experiment_Flow}
\end{figure}

The experimental procedure for each participant is illustrated in Figure~\ref{fig:Experiment_Flow}. Initially, participants completed an FM 100 test on a PC to establish a ground truth regarding the type and severity of CVD. These baseline results were subsequently used for comparison with the CVD assessments conducted in the AR environment. The participants then put on the Meta Quest 3 headset and were given time to familiarize with the device. Next, they were presented with 25 Ishihara plates shown in Figure~\ref{fig:Ishihara-Plate}, and their responses were recorded to evaluate their color discrimination ability. After conducting our adapted FM 100 Hue test on both PC and AR platforms, the results show that all participants have either protan or deutan color vision deficiencies.  Therefore, the Ishihara test was selected as a reference evaluation method due to its efficiency and accuracy. This step provided a baseline measure of color perception in the AR setup without any color correction applied. Subsequently, participants performed the adapted FM 100 test within the AR environment. The results from this test served as input for our personalized recoloring systems to modify participants’ color perception. Finally, participants completed the 25 Ishihara plates again, this time with the personalized color correction enabled. This structured procedure ensured consistent data collection and allowed participants to fully experience the effects of the personalized color correction before providing qualitative feedback.

\section{Quantitative Assessment of Filter Efficacy}
\label{sec:quantitative assessment of filter efficacy}
\subsection{Participant Assessment and Baseline Characterization}

Ten participants volunteered for the experiment, representing various levels of CVD. Table~\ref{tab2} summarizes the results of participants’ color vision assessments and their performance on the Ishihara plate test pre and post applying personalized color correction filter in AR. The CVD type and severity were first determined through a PC-based test, followed by a reassessment within the AR environment. Based on the AR results, severity-personalized color correction filters were applied, and participants then repeated the Ishihara test. The number of correctly identified plates (out of 25) is reported for both the pre- and post-filter conditions. Notably, three participants achieved perfect scores on the desktop FM 100 test yet failed to correctly identify several Ishihara plates without color correction filter. This suggests the presence of mild trichromatic anomalies that were not captured by the FM 100 test but became evident under different testing conditions. These findings highlight the added diagnostic sensitivity of using multiple assessment tools and the practical relevance of personalized AR-based correction, even for individuals with subtle or borderline CVD profiles.

\begin{table}[t]
\caption{Quantitative Results of CVD Assessment and AR-Based Color Correction.}
\label{tab2}
\resizebox{\textwidth}{!}{%
\begin{tabular}{|l|l|l|l|l|l|}
\hline
\textbf{Participant} & \textbf{Desktop CVD} & \textbf{AR CVD} & \textbf{Applied} & \textbf{Pre-Filter} & \textbf{Post-Filter} \\
\textbf{ID} & \textbf{Assessment} & \textbf{Assessment} & \textbf{Filter} & \textbf{Ishihara (/25)} & \textbf{Ishihara (/25)} \\
\hline
1  & Protan, 30\%   & Protan, 50\%   & Protan, 50\%   & 14 & 24 \\
2  & None           & Protan, 30\%          & Protan, 30\%   & 22 & 25 \\
3  & Deutan, 30\%   & Deutan, 30\%   & Deutan, 30\%   & 20 & 24 \\
4  & None           & Protan, 30\%        & Protan, 30\%   & 21 & 25 \\
5  & Protan, 70\%   & Protan, 100\%   & Protan, 100\%  & 12 & 24 \\
6  & Protan, 50\%   & Protan, 50\%   & Protan, 50\%   & 11 & 19 \\
7  & None           & Protan, 20\%   & Protan, 20\%   & 12 & 22 \\
8  & Deutan, 40\%   & Deutan, 40\%   & Deutan, 40\%   & 14 & 23 \\
9  & Protan, 60\%   & Protan, 60\%   & Protan, 60\%   & 11 & 22 \\
10 & Deutan, 80\%   & Deutan, 70\%   & Deutan, 70\%   & 7  & 18 \\
\hline
\end{tabular}
}
\end{table}

\subsection{Statistical Analysis of Color Discrimination Improvements}
\begin{figure}[b]
    \centering
    \includegraphics[width=0.9\linewidth]{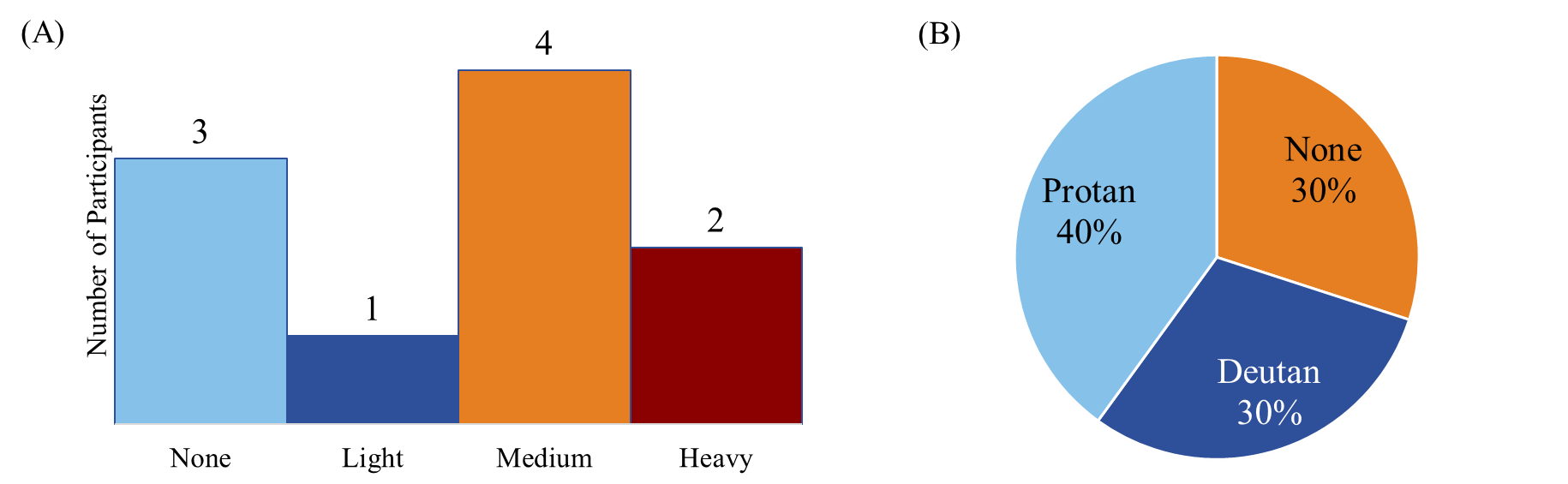}
    \caption{Distribution of Color Blindness Among Participants: (A) Severity Histogram, (B) Type Distribution. Severity is categorized as light (10-40\%), medium (50-80\%).}
    \label{fig:participants}
\end{figure}

The distribution of color blindness type and severity for all participants is presented in Figure~\ref{fig:participants}. The FM 100 Test was administered to all participants to evaluate color discrimination performance before and after filter application.

The paired-samples t-test was selected as the primary statistical method because it is specifically designed to compare two related measurements from the same participants, making it ideal for analyzing with-and-without customized filter. This within-subjects design controls for individual differences in baseline color vision abilities, providing greater statistical power than between-subjects comparisons while reducing the influence of confounding variables. The t-test assumptions were verified through normality testing of the difference scores, and the method's robustness makes it particularly suitable for the relatively small sample size typical in specialized vision research studies.

\textbf{Paired-Samples t-Test Equation:}

\begin{equation}
t = \frac{\bar{d}}{s_d/\sqrt{n}}
\end{equation}

where:
\begin{itemize}
    \item $t$ = t-statistic
    \item $\bar{d}$ = mean of the difference scores (post-filter score - pre-filter score)
    \item $s_d$ = standard deviation of the difference scores
    \item $n$ = number of participants
\end{itemize}

\textbf{Cohen's d Effect Size Equation:}

\begin{equation}
d = \frac{\bar{d}}{s_d}
\end{equation}

where:
\begin{itemize}
    \item $d$ = Cohen's d effect size
    \item $\bar{d}$ = mean of the difference scores
    \item $s_d$ = standard deviation of the difference scores
\end{itemize}

\textbf{Confidence Interval for Mean Difference:}

\begin{equation}
CI = \bar{d} \pm t_{\alpha/2,df} \times \frac{s_d}{\sqrt{n}}
\end{equation}

where:
\begin{itemize}
    \item $CI$ = confidence interval
    \item $t_{\alpha/2,df}$ = critical t-value for the desired confidence level and degrees of freedom
    \item $df = n - 1$ (degrees of freedom)
\end{itemize}

Paired-samples $t$-tests were conducted to evaluate the effectiveness of customized filters on Ishihara scores. Table~\ref{tab:fm100_results} presents the complete statistical analysis for both the overall sample and the subgroup of participants with CVD.
Analysis of the complete sample (N = 10) revealed a statistically significant improvement in FM 100 test scores following filter use with a mean difference of 8.2 points and a large effect size (Cohen's \textit{d} = 2.46). Among the participants with CVD ($n = 7$), this subgroup showed even greater improvement with a mean difference of 9.29 points and a large effect size (Cohen's \textit{d} = 1.45).

\begin{table}[htbp]
\centering
\caption{Paired-samples $t$-test results for Ishihara test scores before and after filter}
\label{tab:fm100_results}

\begingroup
\renewcommand{\arraystretch}{1.5} 
\setlength{\tabcolsep}{12pt} 

\begin{tabular}{lcccccc}
\hline
\textbf{Group} & \textbf{$df$} & \textbf{$t$} & \textbf{95\% CI} & \textbf{$p$} \\
\hline
All Participants        & 9 & 7.79 & [5.82, 10.58] & $<.001$ \\
Participants with CVD   & 6 & 9.13 & [6.80, 11.77] & $<.001$ \\
\hline
\end{tabular}
\endgroup
\end{table}

The results demonstrate that customized filters significantly enhance color vision performance across all participants, with particularly pronounced benefits in individuals with CVD. The large effect sizes indicate that these improvements are not only statistically significant but also practically meaningful. These findings align with our proposal on optical filters that compensate for CVD by selectively modifying spectral composition. The particularly strong effects observed may be attributed to the customized nature of the filters, as individually tailored interventions can address specific spectral sensitivity patterns unique to each participant. While these results offer significant practical implications for individuals with CVD—providing a non-invasive intervention that could improve quality of life in color-dependent activities—the underlying mechanisms require further investigation to determine whether benefits reflect genuine color discrimination enhancement or compensatory visual strategies.

\section{Discussion}
\label{sec:conclusion}
\subsection{Key Contributions and Findings}

This paper presented the development and evaluation of the Hue4U system, a novel approach for personalized color vision CVD detection and correction using consumer-grade AR technology. Our system represents a significant advancement in accessibility by combining the established FM100 test with severity-personalized Daltonization algorithms, creating a practical solution that eliminates the need for prior clinical diagnosis while providing individualized color correction.

The results demonstrate substantial improvements in color discrimination performance following filter application. The large effect sizes (Cohen's $d > 1.4$) indicate statistically significant improvements in CVD correction. These findings are particularly pronounced among participants with CVD, who showed marked enhancement in their ability to distinguish colors that were previously problematic. The magnitude of these effects suggests that personalized filter approaches can meaningfully improve daily visual experiences for individuals with color vision deficiencies.

\subsection{Theoretical and Practical Implications}

Our findings contribute to the growing body of evidence supporting personalized approaches to assistive vision technology. Unlike traditional one-size-fits-all color correction methods, the Hue4U system's ability to adapt to individual severity levels represents a paradigm shift toward precision-based visual aids. This personalization is crucial because CVD manifests with considerable individual variation in both type and severity, necessitating tailored interventions.

From a practical standpoint, the integration with consumer-grade AR devices democratizes access to sophisticated color vision assistance. This approach has significant implications for educational settings, workplace accommodations, and daily life activities where color discrimination is essential. The system's non-invasive, real-time correction capabilities could particularly benefit individuals in color-critical professions or educational contexts.

\subsection{Study Limitations and Future Research Directions}

Several limitations should be acknowledged when interpreting these findings, which also inform important directions for future research. The relatively small sample size ($N = 10$), while appropriate for specialized vision research and sufficient to detect large effects, may limit the generalizability of findings across diverse populations and CVD subtypes. The participant pool's demographic characteristics and the specific types of CVD represented may not fully capture the broader spectrum of color vision deficiencies in the general population. Future investigations should address this limitation by expanding the sample size and diversifying participant demographics to strengthen the generalizability of findings and enable subgroup analyses based on CVD type and severity. Recruiting participants across different age groups, occupational backgrounds, and cultural contexts would provide insights into the system's effectiveness across varied populations.

Additionally, this study relied exclusively on laboratory-based color discrimination testing using the FM100 protocol, which may not fully capture real-world color vision performance. The controlled laboratory environment differs substantially from the complex, variable lighting conditions and diverse color contexts individuals encounter in daily life. To address this limitation, future studies should investigate the system's performance across diverse environmental conditions---including varying lighting, outdoor settings, and complex visual scenes---to provide crucial validation for practical deployment.

The current methodology focuses solely on quantitative pre- and post-test measurements, potentially missing important qualitative aspects of the user experience. Future research would benefit from incorporating mixed-methods approaches that combine quantitative performance measures with qualitative user experience data. Structured interviews and subjective reports could provide valuable insights into user satisfaction, perceived effectiveness, and real-world application challenges. Such qualitative data would complement objective performance metrics and inform system refinements.

Furthermore, longitudinal studies examining the persistence of benefits over extended periods of filter use are essential for understanding the technology's long-term effectiveness. Questions remain about potential adaptation effects, user compliance, and sustained improvement in real-world color vision tasks. These studies would address current gaps in understanding the system's effectiveness beyond immediate laboratory testing and provide critical insights for real-world implementation and long-term user adoption.

\bibliographystyle{splncs04}  
\bibliography{reference}     

\begin{thebibliography}{10}
\providecommand{\url}[1]{\texttt{#1}}
\providecommand{\urlprefix}{URL }
\providecommand{\doi}[1]{https://doi.org/#1}

\bibitem{brettel_computerized_1997}
Brettel, H., Viénot, F., Mollon, J.D.: Computerized simulation of color appearance for dichromats  \textbf{14}(10),  2647--2655. \doi{10.1364/JOSAA.14.002647}, \url{https://opg.optica.org/josaa/abstract.cfm?uri=josaa-14-10-2647}, publisher: Optica Publishing Group

\bibitem{Dalton1794}
Dalton, J.: Extraordinary Facts Relating to the Vision of Colours: With Observations. Self-published, Manchester, England (1794), public domain; Courtesy of Science History Institute

\bibitem{Ishihara1917}
Ishihara, S.: Tests for Colour-Blindness. H. K. Lewis \& Co. Ltd., London, UK (1917), original edition; includes the 38 plates. The 25-plate test is a shortened version commonly used in clinical settings.

\bibitem{OpenXR}
{Khronos Group}: Openxr: Open standard for xr applications. \url{https://www.khronos.org/openxr/} (nd), accessed: 2025-07-17

\bibitem{Arpadffy-Lovas2024}
Árpádffy Lovas, T., Tóth-Molnár, E.: The d15 color arrangement test retains its diagnostic value regardless of display accuracy: A modeling study. medRxiv  (2024). \doi{10.1101/2024.11.15.24314633}, \url{https://doi.org/10.1101/2024.11.15.24314633}, preprint, not peer-reviewed

\bibitem{machado_physiologically-based_2009}
Machado, G.M., Oliveira, M.M., Fernandes, L.A.F.: A physiologically-based model for simulation of color vision deficiency  \textbf{15}(6),  1291--1298. \doi{10.1109/TVCG.2009.113}, \url{https://ieeexplore.ieee.org/document/5290741}

\bibitem{Melillo2017}
Melillo, P., Riccio, D., Di~Perna, L., Sanniti Di~Baja, G., De~Nino, M., Rossi, S., Testa, F., Simonelli, F., Frucci, M.: Wearable improved vision system for color vision deficiency correction. IEEE Journal of Translational Engineering in Health and Medicine  \textbf{5},  3800107 (May 2 2017). \doi{10.1109/JTEHM.2017.2679746}

\bibitem{MetaQuest3}
{Meta Platforms}: Meta quest 3: Advanced all-in-one vr headset. \url{https://www.meta.com/quest/quest-3/} (2023), accessed: 2025-07-17

\bibitem{Simunovic2009}
Simunovic, M.: Color vision deficiency. Eye  \textbf{24}(5),  747--755 (2009). \doi{10.1038/eye.2009.251}, \url{https://www.researchgate.net/publication/38114674_Color_vision_deficiency}

\bibitem{Simunovic2016}
Simunovic, M.P.: Acquired color vision deficiency. Survey of Ophthalmology  \textbf{61}(2),  132--155 (2016). \doi{10.1016/j.survophthal.2015.11.004}, \url{https://doi.org/10.1016/j.survophthal.2015.11.004}, received 1 November 2014, Revised 6 November 2015, Accepted 11 November 2015, Available online 30 November 2015

\bibitem{Tang2020}
Tang, Y., Zhu, Z., Toyoura, M., Go, K., Kashiwagi, K., Fujishiro, I., Mao, X.: {ALCC-Glasses: Arriving Light Chroma Controllable Optical See-Through Head-Mounted Display System for Color Vision Deficiency Compensation}. Applied Sciences  \textbf{10}(7), ~2381 (2020). \doi{10.3390/app10072381}, \url{https://doi.org/10.3390/app10072381}, extended version of paper presented at VRCAI 2018

\bibitem{Tian2022}
Tian, Y., Tang, H., Kang, T., Guo, X., Wang, J., Zang, J.: Inverse-designed aid lenses for precise correction of color vision deficiency. Nano Letters  \textbf{22}(5),  2094--2102 (2022). \doi{10.1021/acs.nanolett.2c00262}, \url{https://doi.org/10.1021/acs.nanolett.2c00262}, epub 2022 Feb 28

\bibitem{verriest_new_1982}
Verriest, G., Laethem, J.V., Uvijls, A.: A new assessment of the normal ranges of the farnsworth-munsell 100-hue test scores  \textbf{93}(5),  635--642. \doi{10.1016/S0002-9394(14)77380-5}, \url{https://www.sciencedirect.com/science/article/pii/S0002939414773805}

\bibitem{Vienot1999}
Viénot, F., Brettel, H., Mollon, J.D.: Digital video colourmaps for checking the legibility of displays by dichromats. Color Research \& Application  \textbf{24}(4),  243--252 (1999). \doi{10.1002/(SICI)1520-6378(199908)24:4<243::AID-COL5>3.0.CO;2-3}, \url{https://doi.org/10.1002/(SICI)1520-6378(199908)24:4<243::AID-COL5>3.0.CO;2-3}

\bibitem{zhou_fast_2024}
Zhou, H., Huang, W., Zhu, Z., Chen, X., Go, K., Mao, X.: Fast image recoloring for red–green anomalous trichromacy with contrast enhancement and naturalness preservation  \textbf{40}(7),  4647--4660. \doi{10.1007/s00371-024-03454-8}, \url{https://doi.org/10.1007/s00371-024-03454-8}

\bibitem{Zhu2022}
Zhu, Z., Toyoura, M., Go, K., Kashiwagi, K., Fujishiro, I., Wong, T.T.: Personalized image recoloring for color vision deficiency compensation. IEEE  (2022), \url{https://ieeexplore.ieee.org/document/9392365}, open Access, under a Creative Commons License

\end{thebibliography}

\end{document}